\documentclass[11pt]{article} 
\usepackage{amsmath}
\usepackage{amsfonts,amssymb,latexsym}

\begin{document}

\newcommand{\nl}{\nonumber\\}
\newcommand{\nnl}{\nl[6mm]}
\newcommand{\nle}{\nl[-2.5mm]\\[-2.5mm]}
\newcommand{\nlb}[1]{\nl[-2.0mm]\label{#1}\\[-2.0mm]}
\newcommand{\ab}{\allowbreak}

\renewcommand{\leq}{\leqslant}              
\renewcommand{\geq}{\geqslant}

\newcommand{\be}{\bes}
\newcommand{\ee}{\ees}
\newcommand{\bes}{\begin{eqnarray}}
\newcommand{\ees}{\end{eqnarray}}
\newcommand{\eens}{\nonumber\end{eqnarray}}

\renewcommand{\/}{\over}
\renewcommand{\d}{\partial}
\newcommand{\no}[1]{{\,:\kern-0.7mm #1\kern-1.2mm:\,}}
 
\newcommand{\al}{\alpha}
\newcommand{\bt}{\beta}
\newcommand{\mm}{{\mathbf m}}
\newcommand{\nn}{{\mathbf n}}

\newcommand{\ZZ}{{\mathbb Z}}

\title{{On the problem of time in two and four dimensions}}

\author{T. A. Larsson \\
Vanadisv\"agen 29, S-113 23 Stockholm, Sweden\\
email: thomas.larsson@hdd.se}

\maketitle 

\begin{abstract} 
In general-covariant theories the Hamiltonian is a constraint, and hence
there is no time evolution; this is the problem of time. In the
subcritical free string, the Hamiltonian ceases to be a constraint after
quantization due to conformal anomalies, and time evolution becomes
non-trivial and unitary. It is argued that the problem of time
in four dimensions can be resolved by a similar mechanism.
This forces us to challenge some widespread beliefs, such as the idea
that every gauge symmetry is a redundancy of the description.
\end{abstract}

\newpage

Time evolution of a quantum-mechanical wave function is given by the
Schr\"odinger equation,
\be
i\hbar{d\Psi\/dt} = H \Psi,
\label{Schr}
\ee
where $H$ is the Hamiltonian. In general-covariant theories,
like general relativity, all spacetime diffeomorphisms are constraints.
In particular, the Hamiltonian is replaced by a Hamiltonian constraint
$H \approx 0$, and the wave function becomes independent of time, 
$\Psi(t) \approx \Psi(0)$. This is the infamous problem of time, which 
has led to much confusion over the years \cite{Ish92}. Although
it is sometimes asserted that it is not a real problem,
the premise of this note is that understanding 
the problem of time is a prerequisite for quantum gravity.

It is clear that an analogous problem will arise in any theory where the
Hamiltonian is a constraint. A well-known example is the free bosonic
string in $D$ dimensions. Whereas pure gravity in two dimensions is a
trivial theory, because the Einstein action is a topological invariant
(the Euler characteristic), the Polyakov action 
\be
S_P = -{1\/2}\int d^2x\ \sqrt{h(x)}\ h^{\al\bt}(x)\ \eta^{\mu\nu}
\ \d_\al\phi_\mu(x)\ \d_\bt\phi_\nu(x)
\label{Pol}
\ee
may be considered as two-dimensional gravity coupled to $D$ massless 
scalars. The Polyakov action depends on a background Minkowski metric 
in target space, but it is background independent on the worldsheet.

The classical theory is invariant under both worldsheet diffeomorphisms
and Weyl transformations. In lightcone quantization, the former are 
gauge-fixed, leaving an infinite conformal symmetry with two
infinite sets of generators $L^R_m$ and $L^L_m$, $m \in \ZZ$. These
generators satisfy two commuting centerless Virasoro algebras,
\be
[L^R_m, L^R_n] = (n-m) L^R_{m+n},
\ee
and similar for $L^L_m$. The Hamiltonian
\be
H = L^R_0 + L^L_0
\label{Ham}
\ee
is classically a constraint, so the problem of time applies here.
Upon quantization, the constraint algebra acquires a central extension,
\be
[L_m, L_n] = (n-m) L_{m+n} - {c\/12}(m^3-m)\delta_{m+n},
\label{Vir}
\ee
where the conformal anomaly $c = 26-D$. According to the no-ghost 
theorem \cite{GSW86}, we can now distinguish three cases.
\begin{itemize}
\item
$D>26$. The Hilbert space has negative-norm states, and thus the
theory is inconsistent.
\item
$D=26$. The unreduced Hilbert space contains null states. The
Weyl symmetry can be factored out because the conformal anomaly
vanishes, and the reduced Hilbert space has a positive-definite
inner product. The Hamiltonian remains a constraint after quantization.
\item
$D<26$. The unreduced Hilbert space has a positive-definite inner
product. It is not possible to pass to a reduced Hilbert space
because of the anomaly, but neither is it necessary to do so, since
already the unreduced Hilbert space is free of ghosts. The Hamiltonian
(\ref{Ham}) is no longer a constraint, but rather the generator of an
ordinary symmetry. Time evolution is non-trivial and unitary, and the
problem of time disappears.
\end{itemize}

That the subcritical free string does not lead to inconsistencies may be
unfamiliar, but it is clearly stated in Chapter 2 of 
\cite{GSW86} that
the free string has a ghost-free spectrum when $D\leq26$; other
string theory textbooks are less clear on this point. It is true that
the subcritical string becomes inconsistent when interactions are added,
and that $D=26$ is special already for the free theory. However, if we
regard the Polyakov action (\ref{Pol}) as a {\em bona fide} theory of 
quantum gravity in two dimensions, it is important to realize that there 
is nothing wrong with $D<26$. Moreover, the special nature of $D=26$ is
less attractive from the viewpoint of this paper; this is the only
value of $D$ where the problem of time persists after quantization. The 
subcritical free string is a perfectly well-defined quantum theory, with 
unitary and non-trivial time evolution, and there is no problem of 
time due to the conformal anomaly.

Two dimensions is one thing, four dimensions is
another. Since the free string is quantum gravity in two dimensions, it
is natural to expect that four-dimensional gravity should be quantized
along similar lines. This was argued by Nicolai et. al \cite{NPZ05}, but
only as a means to discredit Loop Quantum Gravity. The same observation
was made by Jackiw \cite{Jac95b}, with the more constructive goal of
gaining insight into physical gravity in four dimensions; this is also 
the motivation of this note. However, several objections can be 
anticipated.

One objection is that local Weyl scalings of the metric,
\be
h^{\al\bt}(x) \longrightarrow \Lambda(x) h^{\al\bt}(x),
\label{Weyl}
\ee
are peculiar to two dimensions. However, if we choose to gauge-fix 
the Weyl symmetry rather than worldsheet diffeormorphisms, it is
the diffeomorphism symmetry which becomes anomalous
\cite{Jac95a}. Hence we can trade diff and Weyl anomalies in this sense.
Jackiw showed that the number of gauge degrees of freedom which become
physical after quantization is independent of which symmetry we choose
to gauge-fix. 

A second objection is that only in two dimensions is the constraint
algebra of canonical gravity a proper Lie algebra; the relevant algebra
in four dimensions is the Dirac algebra of ADM constraints, 
which is an open algebra
depending on the spatial metric. However, the difference between the
Dirac algebra and the spacetime diffeomorphism algebra is an artefact of
the foliation of spacetime into space and time. It has been noted by many
authors \cite{Ash91,CW87,Rov02} that phase space is a covariant concept;
it is the space of solutions to the equations of motion. Non-covariance
only arises when phase space is coordinatized by its intersection with a
spacelike surface. In any covariant approach, the
constraint algebra of general relativity is the algebra of spacetime
diffeomorphisms in four dimensions. Manifestly covariant canonical
quantization can be performed within the histories approach
\cite{IL95,Sav98,Sav04}, and the closely related
formalism developed in \cite{Lar04a}--\cite{Lar05b}.

A third objection is that in field theory, there are no pure
gravitational anomalies in four dimensions \cite{BPT86}; gravitational
anomalies only exist if spacetime has dimension $4k+2$ \cite{GSW86}.
However, anomalies manifest themselves as non-trivial extensions of
the constraint algebra, and the diffeomorphism algebra in any number of
dimensions certainly admits extensions which generalize the Virasoro
algebra \cite{Lar91,RM94}. In a Fourier basis on the $N$-dimensional
torus, the generators $L_\mu(m) = \exp(im_\rho x^\rho) \d_\mu$
are labelled by momenta $m = (m_\mu) \in \ZZ^N$. It is easy to check
that the following relations define a generalization of the
Virasoro algebra (\ref{Vir}) to $N$ dimensions, apart from a trivial
cocycle.
\bes
[L_\mu(m), L_\nu(n)] &=& n_\mu L_\nu(m+n) - m_\nu L_\mu(m+n) \nl 
&&  - (c_1 m_\nu n_\mu + c_2 m_\mu n_\nu) m_\rho S^\rho(m+n), \nl
{[}L_\mu(m), S^\nu(n)] &=& n_\mu S^\nu(m+n)
 + \delta^\nu_\mu m_\rho S^\rho(m+n), 
\label{mVir} \\
{[}S^\mu(m), S^\nu(n)] &=& 0, \nl
m_\mu S^\mu(m) &\equiv& 0.
\eens
The resolution to this apparent paradox is that we must introduce and
quantize the observer's trajectory\footnote{The clock's worldline 
may be a better name.} $q^\mu(t)$ in addition to the 
fields, because in all representations the extension is given by
\be
S^\mu(m) = {1\/2\pi i} \int_0^{2\pi} 
 dt\ \dot q^\mu(t)\ \exp(im_\rho q^\rho(t)).
\ee
Unless we introduce the trajectory $q^\mu(t)$ in the first place, it is
clearly impossible to write down the relevant anomaly. 

The importance of explicitly introducing the observer was in a sense
anticipated by Rovelli \cite{Rov96}, who made the important observation
that different observers in quantum mechanics may give different
accounts of the same sequence of events. There is a simple physical
motivation for this: the act of observation scatters the observer.
This was explicitly shown for the free scalar field in \cite{Lar04a}.
When acting on a plane-wave state with momentum $k_\mu$, the Hamiltonian
pulls out the operator $k_\mu u^\mu$, where $u^\mu = \dot q^\mu$ is
the observer's four-velocity. If $u^\mu$ is a c-number, this is simply
the Lorentz-invariant form of the energy $k_0$. However, $u^\mu$ is
not a c-number, so rather than just measuring the energy of the 
plane-wave, the Hamiltonian excites an observer quantum. Only in the
limit that the observer is macroscopic and classical (the Copenhagen
limit) is the usual result recovered. Although a classical observer
is often an excellent approximation, we know that all objects are 
fundamentally quantum. The fundamentally incorrect notion of 
a classical observer must be rejected.

A fourth objection is that lowest-energy representations would lead to
anomalies for all gauge theories, not only gravity but also Yang-Mills.
At first sight this conclusion seems absurd, since Yang-Mills theory has
been successfully quantized without anomalies. Nevertheless, this is an
unavoidable consequence if we want to quantize gauge theories in the
same way as we quantize the free string. Nobody can of course deny 
the pragmatic success of path-integral quantization and the
renormalization programme, but despite conceptual clarification brought
about by the renormalization group, it remains to some extent a method
for sweeping infinities under the rug. In contrast, lightcone
quantization of the free string is completely satisfactory from a
mathematical viewpoint, being nothing but part of the representation 
theory of 
the Virasoro algebra. In any event, anomalies inevitably arise from
normal-ordering effects in lowest-energy representations of gauge
algebras \cite{Lar98,RM94}. It should be noted that observer-dependent 
anomalies, being proportional to the second Casimir operator, are 
unrelated to chiral fermion type anomalies proportional to the third 
Casimir.

Consider the spacetime constraint algebra of Yang-Mills theory, with
structure constants $f^{ab}{}_c$. The relevant higher-dimensional
generalization of affine Kac-Moody algebras is, in a Fourier basis,
\be
[J^a(m), J^b(n)] = if^{ab}{}_c J^c(m+n)
+ k \delta^{ab} m_\rho S^\rho(m+n),
\label{gauge}
\ee
where $S^\mu(m)$ is the same as in (\ref{mVir}). Since $q^\mu(t)$
commutes with the gauge generators $J^a(m)$, it can be 
replaced by a c-number. In conventional canonical quantization, it is
natural to assume that the observer is at rest, so 
$q^\mu(t) = \delta^\mu_0 t$. The algebra (\ref{gauge}) becomes 
\be
[J^a(m_0,\mm), J^b(n_0,\nn)] = if^{ab}{}_c J^c(m_0+n_0,\mm+\nn)
+ k \delta^{ab} m_0 \delta_{m_0+n_0},
\ee
where $m = (m_0,\mm)$ is the $3+1$ decomposition of the four-momentum.
Two things are worth noting:
\begin{itemize}
\item
The spatial subalgebra, generated by $J^a(0,\mm)$, is anomaly free.
Hence one may be tempted to pass to a reduced Hilbert space by modding
out spatial gauge transformations. However, doing so is incorrect if
$k\neq0$, because then the full algebra of time-dependent gauge 
transformations, which does act on the kinematical Hilbert space,
is not anomaly free.
\item
The anomalous term does not conserve spatial momentum $\mm$. This is
because we assumed that the observer is at rest, so it can absorb
momentum.
\end{itemize}

A fifth objection is that a gauge symmetry is a mere redundacy of the
description. Both classical and quantum systems can clearly be described
in more or less redundant terms, but there is a minimal 
description where all degrees of freedom are physical. It may happen
that the minimal description of the quantum system has more degrees of
freedom than its classical counterpart. If so, some physical fields 
decouple as gauge fields in the classical limit. Such fake
gauge symmetries must not be eliminated prior to quantization, because
the gauge fields become physical due to gauge anomalies. It is not
straightforward to distinguish between fake and genuine gauge 
symmetries, which persist after quantization,
simply by looking at the classical action; cf. the free string with
$D<26$ and $D=26$.

The main conclusion in \cite{Lar04b,Lar05a,Lar05b} is that interacting
Yang-Mills theory and gravity have such fake gauge symmetries, provided
that the observer's trajectory is included. Since gauge anomalies turn
first-class constraints into second class, the gauge connection has
three and the metric six physical components on the quantum level. The
free Maxwell field is an exception, because the second Casimir vanishes
for the adjoint representation of $U(1)$, and thus the anomalous term
$k=0$ in (\ref{gauge}). A closely related advantage with gauge
anomalies is that a 
non-trivial charge operator can be defined without explicit reference to
the boundary of spacetime.

Finally, a sixth objection is that coordinates have no meaning
in general relativity. However, the key step when
building quantum representations of the diffeomorphism algebra is to
expand all fields in a Taylor series around the observer's trajectory,
{\em viz.}
\be
\phi(x) = \sum_{\mm}\phi_{\mm}(t) (x-q(t))^\mm,
\ee
where $\mm$ is a multi-index \cite{Lar98}. The observer's trajectory
$q^\mu(t)$ is a material object, and so are the Taylor coefficients
$\phi_{\mm}(t)$, which describe the material field $\phi(x)$ in a
neighborhood of the material trajectory. It is straightforward to
phrase at least classical physics in terms of the Taylor data
$\{q^\mu(t), \phi_{\mm}(t)\}$, by making a Taylor expansion of the
equations of motion. In this formulation, the diffeomorphism group
acts on material objects rather than coordinates.

It may be noted that anomalous diffeomorphism symmetry is compatible
with non-trivial correlation functions. In particular, anomalous
dimensions do not depend on the metric structure, but are defined solely
in terms of the differentiable structure, which is a background
structure in general relativity. This is well known in the context of
conformal field theory \cite{FMS96}, which can be regarded as
diffeomorphism-invariant field theory in one complex dimension.

Let us summarize the main lessons from the free string,
formulated in a language applicable to gauge theories in general.
\begin{itemize}
\item
Time-dependent gauge transformations are important. The full
algebra of gauge transformations may have an anomaly, even if the
spatial subalgebra (here generated by $H = L^R_0 + L^L_0$ and 
$L^R_0 - L^L_0$) has not.
\item
Time-dependent gauge generators carry energy. Since the vacuum
state has minimal energy, it must be annihilated by negative-energy
gauge generators (here $L^R_m$ and $L^L_m$ for all $m<0$). 
In other words, the
kinematical Hilbert space carries a non-trivial lowest-energy
representation of the algebra of gauge transformations. This is true
even if anomalies cancel and we can mod out the gauge group after
quantization.
\item
Lowest-energy representations generically give rise to anomalies
(here Virasoro extension = conformal anomaly). This is also true for
algebras of Yang-Mills gauge transformations and diffeomorphisms in
higher dimensions \cite{Lar98,RM94}.
\item
Gauge anomalies do not necessarily lead to inconsistency or violation of
unitarity. We can not pass to some reduced Hilbert space due to the
anomaly, but there is no need to do so, provided that the unreduced
Hilbert space already has positive-definite inner product. A
necessary condition is that the anomalous gauge algebra is represented
unitarily.
\item
In the presence of gauge anomalies, fixing a gauge or restricting
to gauge invariant quantities is wrong. Some or all of the classical
gauge degrees of freedom become physical upon quantization; the quantum
theory has more degrees of freedom than its classical limit.
\item
It higher dimensions it is necessary to include and quantize 
the observer's trajectory (or the clock's worldline), because the 
relevant anomalies are functionals of this trajectory.
\item
The problem of time is resolved if the Hamiltonian ceases to be a
constraint due to quantum anomalies.
\end{itemize}


\begin{thebibliography}{99}

\bibitem{Ash91} A. Ashtekar, L. Bombelli and O. Reula, 
  in {\it Mechanics and Geometry: 200 Years after Lagrange},
  edited by M. Francaviglia, 
  Elsevier (1991).

\bibitem{BPT86} L. Bonora, P. Pasti and M. Tonin,
  {\it The anomaly structure of theories with external gravity},
  J. Math. Phys. {\bf 27} (1986) 2259--2270.

\bibitem{CW87} \v C. Crnkovi\'c and E. Witten,
  in {\it  Newton's tercentenary volume},
  edited by S.W. Hawking and W. Israel,
  Cambridge Univ. Press (1987).

\bibitem{FMS96} P. Di Francesco, P. Mathieu, and D. S\'en\'echal,
  {\it Conformal field theory},
  New York: Springer-Verlag, 1996

\bibitem{GSW86} M.B. Green, J.H. Schwarz and E. Witten,
   {\it Superstring theory, volume I: Introduction},
   Cambridge Univ. Press (1987).

\bibitem{Ish92} C.J. Isham,
  {\it Canonical Quantum Gravity and the Problem of Time},
  {\tt gr-qc/9210011} (1992)

\bibitem{IL95} C.J. Isham and N. Linden,
  {\it Continuous histories and the history group in generalised
  quantum theory},
  J. Math. Phys. {\bf 36} (1995) 5392.

\bibitem{Jac95a} R. Jackiw,
  {\it Another View on Massless Matter-Gravity Fields in Two Dimensions},
  {\tt hep-th/9501016} (1995).

\bibitem{Jac95b} R. Jackiw,
  {\it Two lectures on Two-Dimensional Gravity},
  {\tt gr-qc/9511048} (1995).

\bibitem{Lar91} T.A. Larsson,
  {\it Central and non-central extensions of multi-graded Lie algebras},
  J. Phys. A. {\bf 25} (1992) 1177--1184. 

\bibitem{Lar98} T.A. Larsson,
  {\it Extended diffeomorphism algebras and trajectories in jet space},
  Comm. Math. Phys. {\bf 214} (2000) 469--491.
  {\tt math-ph/9810003}

\bibitem{Lar04a} T.A. Larsson, 
  {\it Manifestly covariant canonical quantization I: the free scalar
  field},
  {\tt hep-th/0411028} (2004)

\bibitem{Lar04b} T.A. Larsson, 
  {\it Manifestly covariant canonical quantization II: Gauge theory
  and anomalies},
  {\tt hep-th/0501043} (2005)

\bibitem{Lar05a} T.A. Larsson,
  {\it Manifestly covariant canonical quantization III: Gravity,
  locality, and diffeomorphism anomalies in four dimensions},
  {\tt hep-th/0504020} (2005).

\bibitem{Lar05b} T.A. Larsson,        
  {\it Manifestly covariant canonical quantization of gravity and
  diffeomorphism anomalies in four dimensions},
  in {\it Focus on Quantum Gravity Research},
  ed David C. Moore,
  Nova Science Publishers, to appear
  (2005).

\bibitem{NPZ05}  H. Nicolai, K. Peeters and M. Zamaklar,
  {\it Loop quantum gravity: an outside view},
  to appear in Class. Quant. Grav,
  {\tt hep-th/0501114} (2005).

\bibitem{RM94} S.E. Rao and R.V. Moody,
  {\it Vertex representations for $N$-toroidal Lie algebras and a
  generalization of the Virasoro algebra},
  Comm. Math. Phys. {\bf 159} (1994) 239--264.

\bibitem{Rov96} C. Rovelli,
  {\it Relational quantum mechanics},
  Int. J. of Theor. Phys. 35 (1996) 1637.
  {\tt quant-ph/9609002}

\bibitem{Rov02} C. Rovelli,
  {\it A note on the foundation of relativistic mechanics.
  II: Covariant hamiltonian general relativity},
  {\tt hep-th/0202079} (2002).

\bibitem{Sav98} K. Savvidou,
  {\it The action operator for continuous-time histories},
  J. Math. Phys. {\bf 40} (1999) 5657-5674.
  {\tt  gr-qc/9811078}

\bibitem{Sav04} N. Savvidou,
  {\it General relativity histories theory},
  Braz. J. Phys. {\bf35} (2005) 307-315.
  {\tt gr-qc/0412059}

\end{thebibliography}
\end{document}